\documentclass[aps,prl,twocolumn]{revtex4}
\usepackage{graphicx}
\usepackage{hyperref}
\usepackage{amsmath}
\usepackage{amsfonts}
\usepackage{color}

\input epsf

\newcommand{\ket}[1]{| {#1} \rangle}
\newcommand{\bra}[1]{\langle {#1} |}

\begin{document}

\title{Spatially dependent electromagnetically induced transparency}

\author{ N.\,Radwell$^1$*, T.\,W.\,Clark$^1$, B.\,Piccirillo$^2$, S.\,M.\,Barnett$^1$\ and \href{http://www.physics.gla.ac.uk/~sfrankea/}{S.\,Franke-Arnold}$^1$} 
\affiliation{$^1$SUPA, School of Physics and Astronomy, University of Glasgow, Glasgow G12 8QQ, UK \\ $^2$ Dipartimento di Fisica, Universit\`{a} di Napoli “Federico II”, Complesso Universitario di Monte S. Angelo, 80126 Napoli, Italy}

\date{\today}

\begin{abstract} 

Recent years have seen vast progress in the generation and detection of
structured light, with potential applications %for
in high capacity optical data
storage and continuous variable quantum technologies.% for continuous variables. 
Here we measure the
transmission of structured light through cold rubidium atoms and observe regions of electromagnetically induced transparency (EIT).  We use q-plates to generate a probe beam with azimuthally varying%added some words on next line
 phase and polarisation structure, and its right and left circular polarisation components provide the probe and control of an EIT transition.
 We observe an azimuthal modulation of the absorption profile that is dictated by the%added phase again
 phase and polarisation structure of the probe laser.  
 Conventional EIT systems do not exhibit phase sensitivity.  
We show, however, that a weak transverse magnetic field closes the EIT transitions, thereby generating phase dependent dark states which in turn lead to phase dependent transparency, in agreement with our measurements.  

\end{abstract}

\maketitle

\paragraph{Introduction:}
The coherent interaction of light with atoms can cause quantum interference between the excitation amplitudes of different optical transitions. 
This can dramatically change the optical response of a medium.  Perhaps the most intriguing example of this is electromagnetically induced transparency (EIT)
\cite{Harris1990,Harris1991Observation}, rendering a medium transparent for resonant probe light when simultaneously exposed to an additional control  beam.  The anomalous dispersion associated with the `transparency window' has been exploited for the generation of slow and stopped light \cite{hau1999light,Turukhin2001}, and related techniques have led to  EIT based quantum memories which can store and retrieve optical information  \cite{Phillips2001, Julsgaard2004, Lvovsky2009optical,Specht2011,Heinze2013}.  

In this letter we report spatially varying optical transparency, achieved by exposing an atomic medium to a single light beam with an azimuthally varying polarisation 
and phase structure \cite{Marrucci2006Optical}.  Such a light mode can be described as `classically entangled' in its polarisation and 
angular position \cite{Spreeuw1998,Toppel2014,Marquardt2014}. If this light is driving a Hanle resonance \cite{Renzoni1997},
 the left and right-handed circular polarisation components  constitute the  probe and control for the
EIT transition respectively, leading to an azimuthal variation of the atomic dynamics.    

It is well established that radially polarised light modes can be focused beyond the diffraction limit \cite{Dorn2003}, % plus more recent
and light with an azimuthal polarisation structure has been proposed for enhanced rotational sensing in so-called photonic gears \cite{DAmbrosio2013}.  The interaction of atoms with phase structured light has been exploited in a variety of experiments, for EIT systems \cite{Pugatch2007,Moretti2009,Veissier2013} as well as for four-wave mixing \cite{Walker2012}.  
Very recently the first  phase-preserving quantum memory has been demonstrated by driving EIT
transitions with light entangled in its polarisation and angular position \cite{Ding2013,Nicolas2014}. These experiments 
rely on the fact that phase-dependent optical information can be stored in EIT coherences.  In contrast, here we generate spatially varying atomic dark states, rendering the atoms transparent to light at specific angular positions.  This self-modulation of the incident light beams effectively converts optical phase information into intensity information.   

It has been shown theoretically that phase-dependent population dynamics require closed linkages between the excitation amplitudes \cite{Buckle1986}.  In our case this is realised by coupling the EIT ground states via a weak transverse magnetic field. 
We provide a theoretical description of the phase-dependent interaction, before presenting the experimental set-up and our results.

%-----------------FIGURE------------------------------
\begin{figure}
\includegraphics[width=\columnwidth]{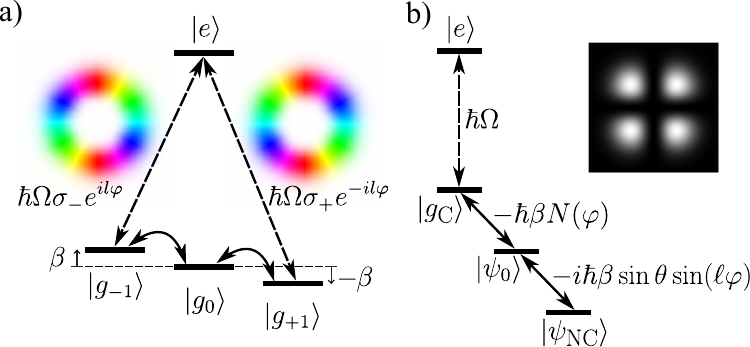}
\caption{\label{fig:1} EIT level scheme with phase dependent driving fields.
	a) Schematic of the atomic energy levels, optical coupling to the excited state (insets show phase represented as hue colour and intensity as saturation), and weak magnetic coupling between the ground states. b) Alternative level scheme expressed in terms of partially
dressed states; the inset is a theoretical simulation of the
absorption profile from $\ket{\psi_{\rm NC}}$ to $\ket{e}$. Details in main text.}
\end{figure}

\vspace{2mm}
\paragraph{Theory:} We consider the atomic $\Lambda$ system shown in Fig.~\ref{fig:1}a).
The atomic ground states, $\ket{g_{\pm 1}}$, are Zeeman sublevels of the same hyperfine state with magnetic quantum numbers $m_{\rm F}=\pm 1$, coupled to the excited state $\ket{e}$  by the  left and right ($\hat{\bf\sigma}_\pm$) circularly polarised components of a probe laser, respectively.  
In the absence of a magnetic field the process is two-photon resonant, providing ideal conditions for EIT. Any detuning from two-photon resonance results in reduced transparency.  Here we are posing the question of whether the transparency of an atomic medium is also affected by a phase difference between the complex excitation amplitudes.  

We consider probe light with an electric field amplitude 
\begin{equation} \label{E} \hat{{\bf E}}(r,\varphi) = \frac{1}{\sqrt{2}} E_0(r) \left( \hat{\bf \sigma}_{+}e^{-i\ell\varphi}+\hat{\bf \sigma}_{-}e^{i\ell\varphi} \right),
\end{equation}
which is correlated in its spin and orbital angular momentum (OAM). Here  $\varphi$ denotes the azimuthal angle and $\ell$ is an integer.  A left circularly polarised photon then has a phase dependence $e^{-i\ell\varphi}$, corresponding to an OAM of $-\ell\hbar$, whereas a right circularly polarised photon has the opposite OAM \cite{franke2008advances,yao2011orbital}.    Each polarisation component has a uniform azimuthal intensity, with a central dark vortex core.  As orthogonal polarisation components do not interfere, the total beam has the same uniform azimuthal intensity.  

This light couples to the atoms via the dipole Hamiltonian  
\begin{align} \label{dipole Hamiltonian}
 \hat{H}_D & =\hat{{\bf D}}\cdot\hat{{\bf E}}=\frac{\hbar\Omega}{\sqrt{2}}\left(e^{-i\ell\varphi}\ket{g_{-1}}\bra{e}+e^{i\ell\varphi}\ket{g_{+1}}\bra{e}\right)+h.c. \nonumber \\
& =\hbar \Omega \ket{g_C}\bra{e} +h.c. 
\end{align}
where $E_0(r)$ is incorporated into the Rabi frequency $\Omega = \Omega(r)$.  The first line is expressed in terms of the atomic levels $\ket{g_{\pm 1}}$, 
and in the second line we have introduced the $\varphi$-dependent partially
dressed states
\begin{equation} \label{C-NC}
\ket{g_{\rm C, NC}}=\frac{1}{\sqrt{2}}\left(e^{-i\ell\varphi}\ket{g_{-1}}\pm e^{i\ell\varphi}\ket{g_{+1}} \right). 
\end{equation}

Note that spontaneous emission causes decay with equal probability into each of the atomic ground states, so that an atom is driven out of the coupling state $\ket{g_{\rm C}}$ into the non-coupling state $\ket{g_{\rm NC}}$ or the unperturbed atomic state $\ket{g_0}$ after very few absorption-emission cycles.  This system exhibits phase-dependent coherences and 
could be employed for EIT based quantum memories for phase-encoded light. 
However, as $\ket{g_{\rm NC}}$ is rotationally symmetric, neither the atomic populations nor the absorption show phase dependence.  This, of course, was expected, as phase-dependent populations require a closed loop level system, where atomic transitions combine interferometrically \cite{Buckle1986,Kosachiov1992}.

A magnetic field, as long as it is not aligned with the probe beam propagation direction, will perturb the rotational symmetry.     
We consider an arbitrary magnetic field ${\bf B}= B (\cos\theta \hat{\bf z}+\sin\theta \hat{\bf x})$, where the light propagates along $\hat{\bf z}$ and for simplicity we have chosen $\hat{\bf x}$ to be the transverse direction of the B-field.

The total interaction Hamiltonian \begin{equation} \label{H} \hat{H}=\hat{H}_{\rm D}+\hat{H}_{\rm B} \end{equation}
 includes, in addition to (\ref{dipole Hamiltonian}), the magnetic interaction Hamiltonian, which in the weak field limit is:
\begin{align}
\hat{H}_{\rm B} & =g_F \mu_B \hat{\bf F}\cdot\vec{\bf B}  \nonumber \\
& =\hbar  \beta    \Big[ \cos \theta (\ket{g_{+1}} \bra{g_{+1}}-\ket{g_{-1}} \bra{g_{-1}}) \Big]   \\  
& \hspace{3mm} -\hbar \beta \Big[ \frac{\sin\theta}{2}(\ket{g_0}\bra{g_{+1}}+\ket{g_0}\bra{g_{-1}} +h.c.) \Big], \nonumber 
\end{align}
where $\hat{\bf F}$ is the total angular momentum operator and we have defined the magnetic parameter $\beta= g_{\rm F} \mu_{\rm B} B $.
The first term describes the Zeeman shift of $\ket{g_{\pm 1}}$ due to $B_{\rm z}$, whereas the second term describes the mixing of $\ket{g_0}$ with $\ket{g_{\pm 1}}$, as illustrated in Fig.~\ref{fig:1}a).  
It is instructive to introduce a modified basis set of partially dressed states, combining $\ket{g_{\rm NC}}$ and $\ket{g_0}$: 
\begin{align}
\ket{\psi_{\rm 0}}& =\frac{1}{N(\varphi)} \big[ -\cos{\theta} \ket{g_{\rm NC}}+\sin{\theta} \cos{(\ell \varphi)} \ket{g_0} \big],\nonumber \\
\ket{\psi_{\rm NC}}& =\frac{1}{N(\varphi)} \big[\sin{\theta} \cos{(\ell \varphi)} \ket{g_{\rm NC}}+\cos{\theta} \ket{g_0}\big], 
\end{align}
where $N(\varphi)=\sqrt{1-\sin^2\theta\sin^2(\ell\varphi) }$ ensures normalisation.  Together with $\ket{g_{\rm C}}$ these form a complete basis set, in which the Hamiltonian (\ref{H}) can be rewritten as
\begin{align}
\hat{H}=&\hbar\Omega\ket{g_{\rm C}}\bra{e}-\hbar \beta  N(\varphi) \ket{\psi_0}\bra{g_{\rm C}} \\
&-i\hbar\beta \sin\theta \sin(\ell \varphi) \ket{\psi_{\rm NC}}\bra{\psi_{0}} +h.c., \nonumber
\end{align}
forming the ladder system shown in Fig.~\ref{fig:1}b).  In this basis $\ket{g_{\rm C}}$ couples optically to the excitated state and magnetically to $\ket{\psi_{\rm C}}$, which in turn couples magnetically to $\ket{\psi_{\rm NC}}$.  Importantly, at certain angles $\varphi_n=n\pi/\ell$ ($n \in {\mathbb N}$), the state $\ket{\psi_{\rm NC}}$ decouples completely from the optical and magnetic transitions. At those angles, all atoms will spontaneously decay into the dark state $\ket{\psi_{\rm NC}}$ and light can pass unhindered through the atoms. 
The absorption profile can be calculated by evaluating Fermi's Golden rule, $T_{i\to f}\propto \frac{2\pi}{\hbar} |\bra{i}H\ket{f}|^2 $:
\begin{align} \label{2ell}
T_{\psi_{\rm NC}\to e}&\propto\left( \frac{2\pi}{\hbar}\right)^3 |\hbar \Omega|^2 |\hbar \beta|^4  \left|\sin\theta \sin(\ell \varphi) \right|^2 N(\varphi)^2   \nonumber \\
&\xrightarrow{\theta \ll \pi/2}  \left( \frac{2\pi}{\hbar}\right)^3 |\hbar \Omega|^2 |\hbar \beta|^4  |\theta|^2  \left|\sin(\ell \varphi) \right|^2  .
\end{align}

For magnetic fields that
have a small transverse component to the probe light propagation, we find a
sinusoidal variation of the absorption profile with a periodicity of $2 \ell$,
shown in the inset in Fig.~\ref{fig:1}b), whereas for larger $\theta$ the
absorption acquires additional structure with twice the periodicity. We will explore the dependence on the magnetic field direction in more detail in a future paper.

\vspace{3mm}
\paragraph{Experimental Setup \& Procedure:}
%-----------------FIGURE------------------------------
\begin{figure}
\includegraphics[width=\columnwidth]{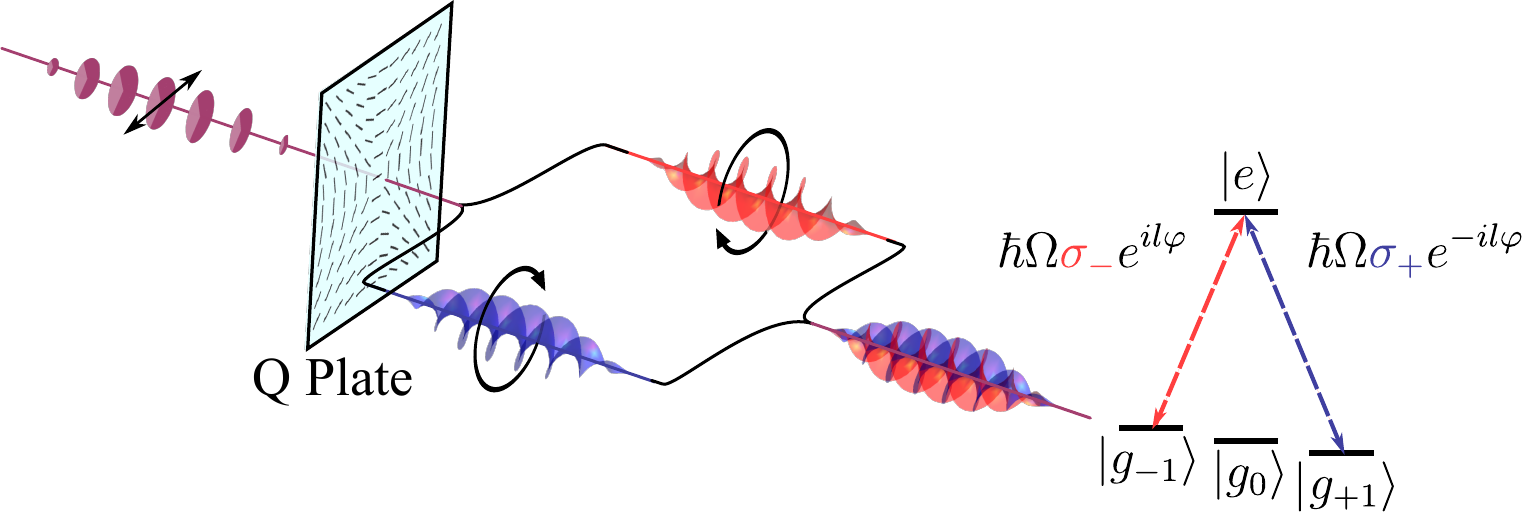}
\caption{\label{fig:2} Probe light generation and differential driving of the atomic transitions.  A linearly polarised Gaussian probe beam impinges onto a q-plate, here shown for $q=1$. Its right and left circular polarisation components are encoded with opposite OAM, driving transition amplitudes from the different magnetic sub-levels of the atoms.}
\end{figure}

The  electric field (\ref{E}) is generated by passing a linearly polarised probe laser through a q-plate \cite{Marrucci2006Optical} with $q=\ell/2$.

Q-plates are liquid-crystal based retardation waveplates with an inhomogeneous optical axis,  exhibiting an azimuthal topological charge $q$. Applying an AC external voltage sets their retardation to $\pi$ \cite{Piccirillo2010} so that they convert left to right handed polarisation, adding an OAM of $2 q \hbar$, and vice versa, i.e.  $\sigma_{\pm}\to \sigma_{\mp} \exp(\pm i 2 q \varphi)$. Consequently, linear polarised light generates classical entanglement between polarisation and azimuthal angle of the probe laser, illustrated in  Fig.~\ref{fig:2}.  

Our experiments are performed on cold, trapped $^{87}$Rb atoms, using the hyperfine transition $5^2S_{1/2} (F=1)\to5^2P_{3/2}( F'=0)$.  We prepare the sample in a
dynamic dark spontaneous force optical trap (SPOT) \cite{Radwell2013},% loaded from a standard magneto-optical trap.
%While our dark SPOT was developed to generate high atomic densities, its main advantage for the work reported here is that the atoms are confined to
providing a high density ($2\times10^{11}{\rm cm}^{-1}$) cloud in the lower  $F=1$ ground state, matching our theoretical model. 

\begin{figure}
\includegraphics[width=\columnwidth]{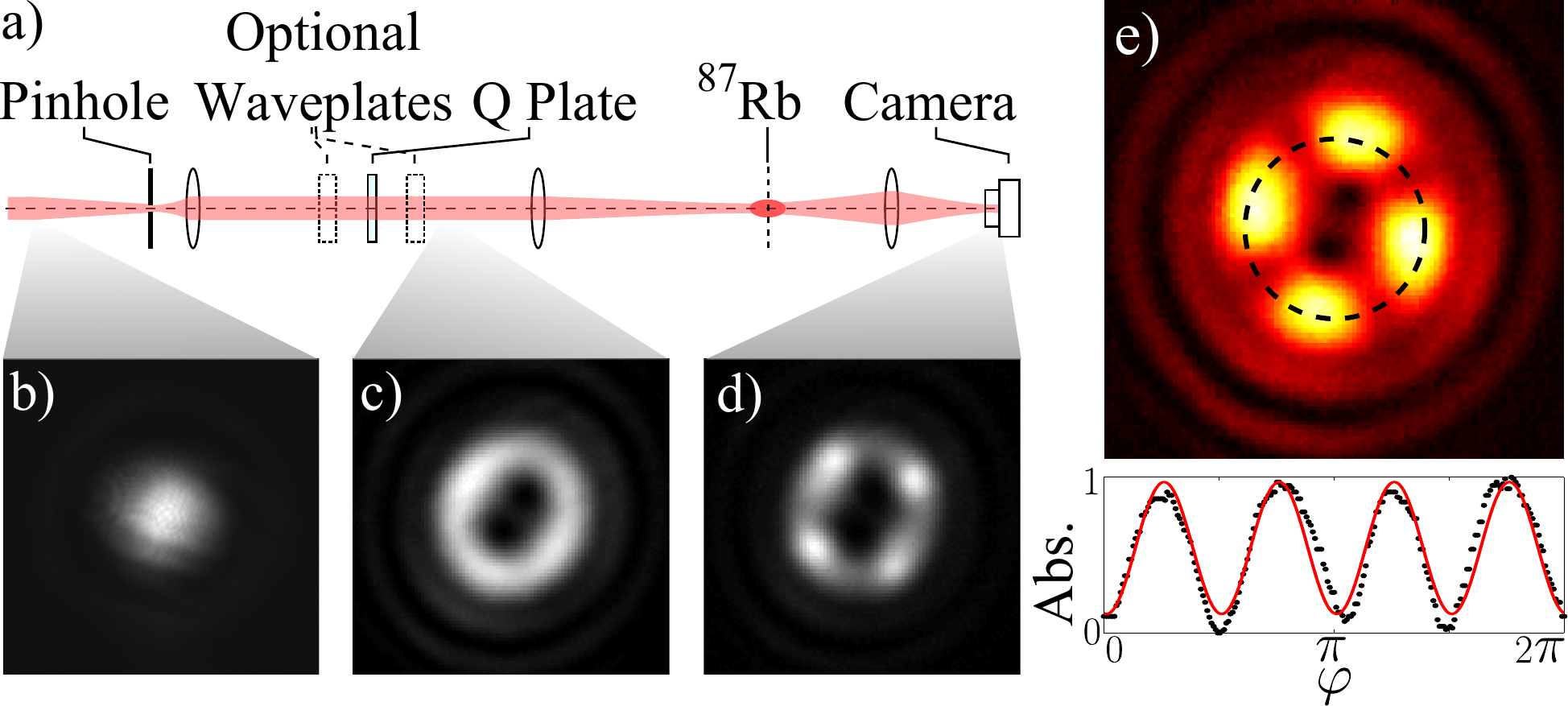}
\caption{\label{fig:3} Experimental setup and initial results. a) Experimental beam path. b) Intensity profile of the input beam, c) of the probe beam generated by the q-plate of $q=1$, and d) after absorption from the atoms.  e) Corresponding absorption image and polar plot at radius of maximal contrast.}
\end{figure}

The  linearly polarised probe beam is spatially filtered and collimated before being passed through the q-plate path, as shown in Fig.~\ref{fig:3}a). Panels b) and
c) show the intensity profile of the probe beam before and after a $q=1$ q-plate. Optional waveplates may be added to alter the polarisation profile. 
A lens  images the far field of the q-plate onto the atoms which are further imaged onto a CCD camera.

An experimental run loads a standard MOT for 6{\,}s, followed by a SPOT loading time of 250{\,}ms.  All trapping lasers
are then switched off and the cloud expands for typically 3~ms to achieve the desired density of $2\times10^{11}{\rm cm}^{-1}$,
chosen to produce the highest contrast absorption images. The q-plate beam is then shone through the atoms for $\sim1\,$ms and images are recorded in the presence of atoms ($I_{\rm Atoms}$), in the absence of atoms ($I_{\rm Probe}$) and without  lasers ($I_{\rm Dark}$).

%-----------------FIGURE------------------------------

During the MOT and SPOT loading the atoms are exposed to the typical quadrupole magnetic field generated by anti-Helmhotz coils, while stray fields are compensated by three orthogonal Helmholtz coils.  During the atomic expansion we switch off the quadrupole field and add a weak linear magnetic field of 0.1$\,$G by modifying the currents in the compensation coils. 

The transparency profile can be observed directly in the transmitted probe intensity (Fig.\ref{fig:3}d)), but in order to quantify the expected sinusoidal absorption profile we present our data in form of absorption images. 
The absorption coefficient is proportional to the optical density, $ {\rm OD}= \log \left(\frac{I_{\rm Probe} - I_{\rm Dark}}{I_{\rm Atoms}-I_{\rm Dark}}\right).$
For beams with low intensity regions the optical density is noisy and/or undefined where the denominator is zero or negative.  We therefore scale the results by a factor of $\sqrt{I_{\rm Probe}}$ in order to enhance the meaningful areas of interest (Fig.~\ref{fig:3}e)). This scaling of course does not affect the azimuthal sinusoidal variation but only the radial intensity profile.  Note that bright areas in the absorption image correspond to  positions of low light transmission and vice versa.

\vspace{3mm}
\paragraph{Results \& Discussion:}
%-----------------FIGURE------------------------------
\begin{figure}
\includegraphics[width=\columnwidth]{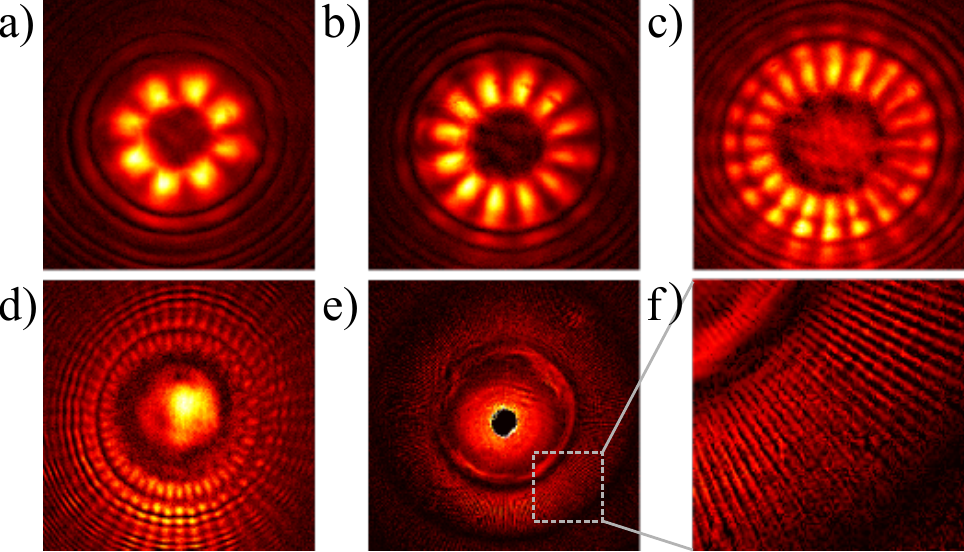}
\caption{\label{fig:4} Absorption patterns for higher q-values, each showing a $4q$-fold symmetry, for $q=2$ (a), $ q=3$ (b), $q=5$ (c), $q=12$ (d) and $q=100$ (e), with a zoom of the marked section in (f).}
\end{figure}

The procedure outlined above generates a probe beam containing right and left handed circularly polarised components 
with $\pm 2\hbar$ units of OAM, which drive transitions from the $m_F=\mp1$ ground states, respectively.
The absorption profile, shown in Fig.~\ref{fig:3}e), shows four-fold symmetry, satisfying the predicted $2\ell$ sinusoidal absorption profile predicted in (\ref{2ell}).

We demonstrate that this result holds also for higher rotational symmetries by using probe beams generated from various different q-plates.% A q-plate of charge $q$ drives  transitions from the $m_F=\mp1$ ground states with light that has an azimuthal phase dependence of $\pm 2 q \phi$ respectively.  
 All observed absorption profiles, shown in Fig.~\ref{fig:4}, clearly display the expected $2\ell$ lobes.

So far we have interpreted spatial transparency as a result of interference between transitions driven by the oppositely-phased right and left circular polarisation components.  Alternatively, we may consider the overall probe polarisation profile.  All linear polarisations can be decomposed into two circular components of equal amplitude, with their relative phase governing the polarisation angle. The probe beams considered for Figs.~\ref{fig:3} and \ref{fig:4} have an azimuthally varying phase difference, corresponding to vector vortex beams.  The excitation amplitudes interfere only at positions where the linear polarisation is aligned with the atom quantisation axis as set by the magnetic field, causing transparency.  

\begin{figure}
\includegraphics[width=\columnwidth]{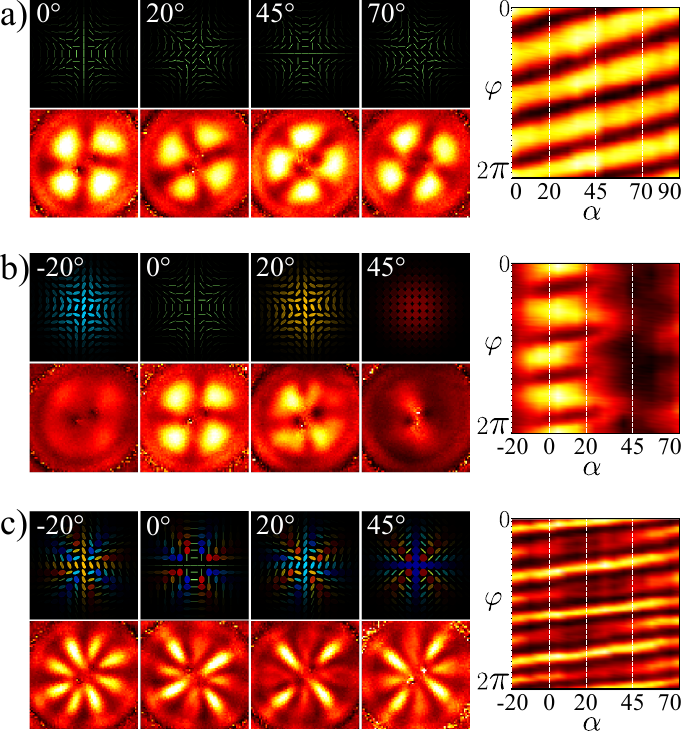}
\caption{\label{fig:5} Absorption profiles as function of  input polarisation for a probe beam with $q=1$. a) Rotation of linear input polarisation via a $\lambda/2$ plate before the q-plate at angle $\alpha$ %with respect to the initial polarisation.  
Probe polarisation profile (top row), and associated measured absorption profile (lower row). To visualise the rotation of the absorption profile with input polarisation we display at the right the polar plot of the absorption profile as a function of $\alpha$,  taken every 5 degrees.  b)  Variation of input ellipticity via a $\lambda/4$ plate infront of the q-plate.  The resulting polarisation profiles vary between a vector vortex beam at $\alpha=0$, associated with maximum contrast in the absorption profile, and a purely circularly polarised beam at $\alpha=45^\circ$, generating minimal contrast. c) Adding a $\lambda/4$ plate after the q-plate changes the symmetry of the observed pattern.}
\end{figure}

Following from this argument, we further probe the EIT mechanism by changing the incident polarisation. Based on the system symmetry, one would expect that rotating the linear polarisation before the q-plate would simply turn the absorption profile. We test this experimentally by rotating a $\lambda/2$-plate placed before the q-plate.  
The observed absorption profiles are shown in Fig.~\ref{fig:5}a) for a few angles of the $\lambda/2$- plate. Their overall rotation, however, can perhaps be seen more clearly from the unwrapped azimuthal profiles displayed on the right panel of Fig.~\ref{fig:5}a) which combines azimuthal data from 19 individual $\lambda/2$-plate positions, confirming that the absorption profile follows the input polarisation. 

We perform a similar experiment with a $\lambda/4$-plate placed before the q-plate.
%Depending on the angle of the $\lambda/4$-plate with respect to the initial linear polarisation, we thus generate an input beam
The relative angle of the input polarisation and the $\lambda/4$-plate governs the ellipticity of the polarisation, ranging from linear to circular.  
For linear polarisation we of course retrieve the patterned absorption as before. Purely circular light, however, will be converted by the q-plate into oppositely circular light and acquire OAM.
It will drive only one arm of the $\Lambda$ transition and, due to the level structure shown in fig.~\ref{fig:1}a), the atoms will be driven into a dark state, namely the electronic levels $\ket{g_{-1}}$ or $\ket{g_{+1}},$ which unlike the state $\ket{\psi_{\rm NC}}$ are spatially uniform.    
The resulting absorption images, as shown in Fig.~\ref{fig:5}b), reveal a clear four-lobed pattern at 0 degrees and uniform
absorption at about 45 degrees.  Unwrapping the data, we obtain the image on the right of Fig.~\ref{fig:5}b), showing the emergence of the absorption structure as a function of the $\lambda/4$-plate angle.  

Fig.~\ref{fig:5}c), finally, shows the increase of the rotational frequency by placing a $\lambda/4$-plate after the q-plate, resulting in $4\ell$ absorption lobes.  
We note that the atomic system essentially behaves as an `inverse' linear polariser, with its axis coincident with the quantization axis set by the magnetic field. The petal-like absorption pattern is reminiscent of the petal mode transmission patterns observed by inserting the q-plate between crossed linear polarizers - our `atomic polariser' however maintains a memory of the incident light. %Of particular  interest is the lower panel in Fig 5 c), where the $\lambda/4$ after the q-plate combined with the linear atomic polarizer (in negative/absorbing version) producing patterns symmetrically identical to those observed with both coherent and incoherent light under microscope.

\vspace{3mm}

\paragraph{Conclusion:}
We have demonstrated spatially structured transparency by probing atoms with vector vortex light. The atomic absorption profile reflects the spatial polarisation variation across the probe beam.  We have shown that structured transparency arises from interference between two excitation amplitudes driven by light with different spatial phase profiles, in combination with weak magnetic coupling between the Zeeman ground states.  The system contains a spatially varying dark state, whose population in turn resulted in a self-modulation of the incident light beams.  Spatial absorption patterns were recorded for a variety of q-plates and input polarisations, showing that the symmetry of the absorption profile is linked to the symmetry of the input polarisation pattern, whereas the contrast of the absorption profile depends on the balance between the excitation amplitudes.  

While we have demonstrated spatially dependent EIT for the special case of vortex light beams, the mechanism applies to any light where the circular polarisation components have a different spatial phase profile.  Furthermore, while we have used classical light fields, in principle even the circular polarisation components of a single photon could be encoded with different phase profiles, and written into atomic dark states.  We hence expect that spatially dependent EIT has applications for the storage of high-dimensional optical information in phase-dependent quantum memories.  Finally we note, that the `classical entanglement' between polarisation and azimuthal angle of the probe laser is converted, via the Hanle resonance, into a non-local correlation between the spatial intensity profile of the transmitted light and the atomic population profile, offering perphaps a vehicle to convert single photon classical entanglement into true quantum entanglement, an effect that we plan to investigate in more detail in future.   
\vspace{3mm}
\paragraph{Acknowledgements:}

We acknowledge the financial support given by the Leverhulme Trust via the project RPG-2013-386, EPSRC programme grant COAM EP/I012451/1 and for the early work by  the European Commission via the FET Open grant agreement Phorbitech FP7-ICT-255914. We thank Lorenzo Marrucci, Fabio Sciarrino, Vincenzo D'Ambrosio and Sergei Slussarenko for useful discussions and above all for the generous provision of various q-plates.  

%\bibliographystyle{apsrev4-1}
%\bibliography{SpatialEIT}


\end{document}